# Experimental observations indicating the topological nature of the edge states on HfTe$_5$

Rui-Zhe Liu(刘睿哲)[1,2,3*], Xiong Huang(黄雄)[1,3*], Ling-Xiao Zhao(赵凌霄)[1,3*], Li-Min Liu(刘立民)[13], Jia-Xin Yin(殷嘉鑫)[4], Rui Wu(武睿)[1,2], Gen-Fu Chen(陈根富)[13,5], Zi-Qiang Wang(汪自强)[6], Shuheng H. Pan(潘庶亨)[1,2,3,5,7**]

[1] Beijing National Laboratory for Condensed Matter Physics, Institute of Physics, Chinese Academy of Sciences, Beijing 100190, China

[2] Physical Science Laboratory, Huairou National Comprehensive Science Center, Huairou, Beijing 101400, China

[3] School of Physics, University of Chinese Academy of Sciences, Beijing 100190, China

[4] Laboratory for Topological Quantum Matter and Advanced Spectroscopy (B7), Department of Physics, Princeton University, Princeton, NJ, USA.

[5] Songshan Lake Material Laboratory, Dongguan, Guangdong 523808, China

[6] Department of Physics, Boston College, Chestnut Hill, Massachusetts 02467, USA.

[7] CAS Center for Excellence in Topological Quantum Computation, University of Chinese Academy of Sciences, Beijing 100190, China

*These authors contributed equally in this work.

**To whom correspondence should be addressed. Email: span@iphy.edu.cn

Email: rzliu@iphy.edu.cn; span@iphy.edu.cn; 297031216@qq.com

**Abstract** The topological edge states of two-dimensional topological insulators with large energy gap furnish ideal conduction channels for dissipationless current transport. Transition metal tellurides XTe$_5$ (X=Zr, Hf) are theoretically predicted to be large-gap two-dimensional topological insulators and the experimental observations of their bulk insulating gap and in-gap edge states have been reported, but the topological nature of these edge states still remains to be further elucidated. Here, we report our low temperature scanning tunneling microscopy/spectroscopy study on single crystals of $HfTe_5$. We demonstrate a full energy gap of ~80 meV near the Fermi level on the surface monolayer of $HfTe_5$ and that such insulating energy gap gets filled with finite energy states when measured at the monolayer step edges. Remarkably, such states are absent at the edges of a narrow monolayer strip of one-unit-cell in width but persist at both step edges of a unit-cell wide monolayer groove. These experimental observations strongly indicate that the edge states of $HfTe_5$ monolayers are not trivially caused by translational symmetry breaking, instead they are topological in nature protected by the 2D nontrivial bulk properties.

A two-dimensional topological insulator (2D TI) is characterized by a full energy gap in the bulk electronic bands and conducting helical states at the one-dimensional (1D) edges. There, backscattering is forbidden because the electrons with opposite spins propagate in opposite directions. Therefore, 2D TIs are proposed to be used in fabrication of dissipationless electronic devices. As a high-resolution local probe, scanning tunneling microscopy/spectroscopy (STM/S) technique enables us to study such promising 2D electronic system and its localized 1D edge states on a microscopic scale. In many proposed 2D topological systems[1-11], e.g. Bi[12], $Bi_{14}Rh_3I_9$[13], $WTe_2$[14, 15], (Pb, Sn)Se[16] and $XTe_5$(X=Zr, Hf) [17-19], STM/S results have already shown the insulating bulk energy gap and the existence of edge states. However, the discussions of the topological nature of these edge states observed by STM/S merely rely on the predictions of theoretical calculations. It would be much preferred if there are further experimental evidences closely related to their topological origin. In this letter, we report our STM/S study on the electronic structures of the surface of single crystal $HfTe_5$. Particularly, we emphasize our investigations of the monolayer step edge,   i.e. a 1D boundary of a 2D bulk, and the interaction of the electronic states of two step edges with a one-unit-cell separation.

We discuss the implications of our experimental observation and argue for the topological nature of the in-gap edge states.

Single crystals of HfTe$_5$ are grown by chemical vapor transport method. Stoichiometric amounts of Hf (powder, 3N, Zr nominal 3%) and Te (powder, 5N) are sealed in a quartz ampoule with iodine (7 mg/mL) and placed in a two-zone furnace. Typical temperature gradient from 500 °C to 400 °C is applied. After one month, long ribbon-shaped single crystals are obtained. To obtain a high-quality surface for STM/S measurements, the HfTe$_5$ samples are mechanically cleaved in situ at ~20 K, and inserted into the STM head immediately after cleaving. All STM measurements are performed at 4.3 K with tungsten tips fabricated by electrochemical etching followed by field emission against a single crystal gold target. STM topography are measured in the constant current mode and differential tunneling conductance spectra were obtained using the lock-in technique.

As shown in Fig. 1(a), the HfTe$_5$ crystal is structured by stacking 2D HfTe$_5$ layers along b axis with a spacing constant of 1.45 nm, and the HfTe$_5$ layer is composed of HfTe$_3$ prismatic chains along the a-axis which is bonded in the c-axis direction by the tellurium zigzag chains. It has an orthorhombic layered structure with the space group of Cmcm. The lattice constants in the a-c plane are a=0.39 nm and c=1.37 nm. Since the coupling between the two adjacent HfTe$_5$ layers is van der Waals type, the cleaving process always takes place there and exposes the HfTe5 layer. A STM topographic image of the HfTe$_5$ layer is displayed in Fig. 1(b). It demonstrates a stripy structure with a periodicity of ~1.40 nm, which is consistent with the value of lattice constant c. The zoom-in topography in Fig. 1(b) clearly shows the topmost Te dimers of the prismatic HfTe$_3$ chains. The measured ~0.40 nm distance between two adjacent Te dimers agrees with the lattice constant a. In Fig. 1(c), we display the differential tunneling conductance spectrum taken on this surface that clearly shows an energy gap of ~80 meV near the Fermi level (E$_F$). The profile of this STS curve resembles that of ZrTe$_5$[17], demonstrating these two materials have very similar electronic structure, as predicted by theoretical calculations[20].

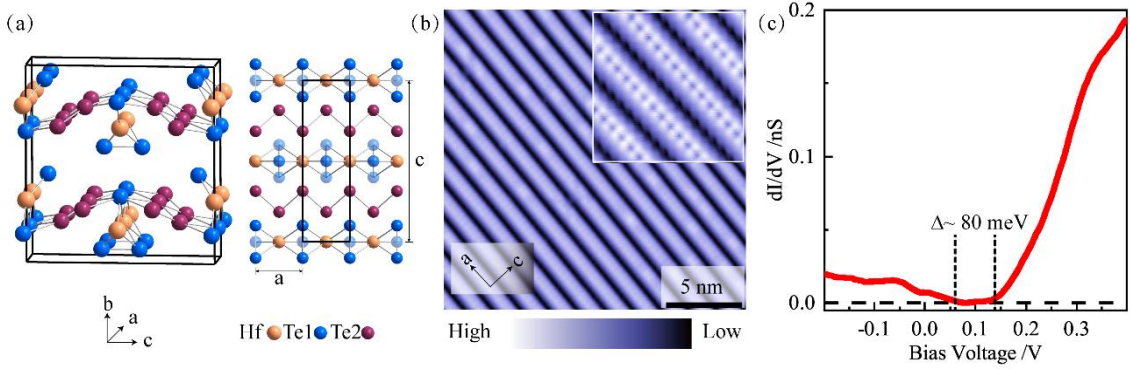

FIG. 1 (a) Side view (left) and top (right) view of the schematic crystal structure of HfTe$_5$. (b) Topographic image of the surface of a cleaved single crystal HfTe$_5$ (20 × 20 nm$^2$, V = 500 mV, I = 0.03 nA). Inset: Zoom-in image of the HfTe$_3$ chains (V = 500 mV, I = 0.1 nA). (c) Tunneling conductance on HfTe$_5$ surface (V = 500 mV, I = 0.5 nA).

Because of the relatively weak coupling between the adjacent HfTe$_3$ prismatic chains, the in-plain cleavage can take place between the chains. Thus, it is natural to expect for the occurrence of straight monolayer steps along the chain direction. Fig. 2(a) shows a topographic image of such a step edge with a height of ~0.70 nm, which equals half of the lattice constant b, indicating that this is a monolayer step edge. A sequence of tunneling spectra taken along a line perpendicularly across the step edge is listed in Fig. 2(b). For those spectra taken at the locations near the step edge, finite density of states (DOS) emerges inside the entire energy gap. To demonstrate the spatial evolution of the edge states from the edge into the 2D bulk, we show in Fig. 2(c) the values of the integrated DOS (within the energy range of 100meV~130meV) for the corresponding spectra in Fig. 2(b). The in-gap states exhibit an exponential decay with a characteristic length $r_0$ ~2.91 nm, twice of the lattice constant c (1.40 nm), demonstrating that the edge states are localized at the step edge and decay into the 2D bulk for a short distance of only a few unit-cells. All these experimental observations are very similar to that on ZrTe$_5$ reported previously[17], and also consistent with the early theoretical predictions that suggest the bulk of HfTe$_5$/ZrTe$_5$ is a weak 3D TI and its monolayer is a possible candidate for 2D topological insulator supporting topological edge states[20].

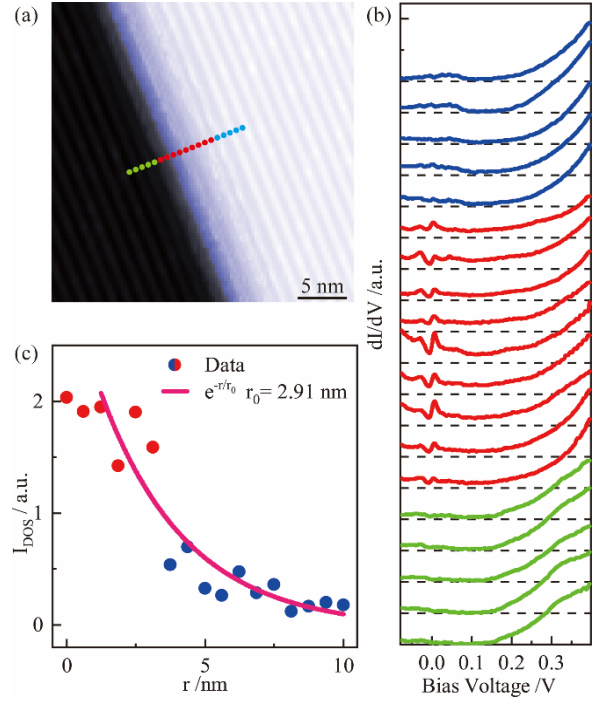

FIG. 2 (a) Topographic image of a monolayer step (20 × 20 nm$^2$, V = 500 mV, I = 0.03 nA). (b) Spatially resolved tunneling spectra (V = 500 mV, I =0.5 nA) across the step edge. Their corresponding locations are marked in (a) as colored dots. (c) Integrated conductance within the gap plotted as a function of distance away from the step edge on the upper terrace surface. The red curve is the exponential fitting with a decay length of ~2.91 nm.

In general, the edge states of a 2D system can be either trivial ones caused by symmetry breaking or nontrivial ones protected by the 2D topological band structure. To further identify the topological nature of the edge states observed at the edges of the HfTe$_5$/ZrTe$_5$ monolayers, more experimental evidences would be much preferred. Here, we present a case of an isolated narrow HfTe$_5$ strip, as shown in Fig. 3(a). This narrow strip resides on the atomically flat surface of the single crystal HfTe$_5$ and has a height of ~0.70 nm and width of one unit-cell (~1.40 nm). From the high resolution STM topography (inset of Fig. 3(a)), we can resolve the atomic structure of this narrow strip as the same of a unit-cell strip from the striped surface lying below. Notably, the spatially resolved tunneling conductance spectra in Fig. 3(b) show that the energy gap keeps open across this strip and no conducting in-gap states emerge at either its edges. Fig. 3(c) compares the "regular" tunneling spectra taken at the edge of a large 2D monolayer, at the edge of a unit-cell strip, and on the monolayer away from the edge.

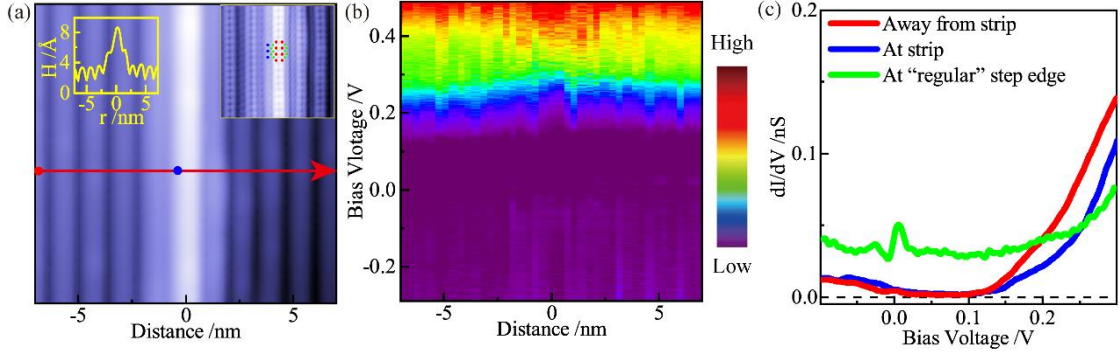

FIG. 3 (a) Topographic image showing an isolated strip (V = 500 mV, I = 0.03 nA). (b) Spatially resolved tunneling spectroscopic map across the strip of red arrow line in (a) showing the absence of edge states (V=500 mV, I =0.5 nA). (c) Tunneling spectra taken at (blue) and away (red) from the strip, respectively. The spectrum with edge states (green) obtained in Fig 2 (c) is presented as a reference.

The absence of in-gap states at the edges of a one-unit-cell wide monolayer strip is quite remarkable. Crystal structurally, there is no difference between the edge of a one-unit-cell monolayer strip and the one of a large monolayer. The trivial edge states due to the lateral environment change at the step edge should appear in both cases. However, it is more consistent to argue that there is a fundamental difference between the two cases for the edge states of topological nature. In the case of the large monolayer, the topological edge states are supported and protected by the topological bands of the 2D topological insulator. While in the case of the one-unit-cell monolayer strip, no such bands exist due to the lack of 2D translational invariance. It can also be argued that the topological edge states from the two edges of a narrow topological strip can couple, when they are close enough, and open a gap. Indeed, in our observation, the two edges are distant from each other for only a unit-cell length, while the characteristic decay length of the edge states is several unit-cells long, as shown in Fig. 2(c). Therefore, the strong coupling between the topological edge states opens a full gap as the one on a large 2D topological monolayer.

The topological argument can also be further supported by the experimental observation on a one-unit-cell wide monolayer groove (Fig. 4(a)). As shown in Fig.

4(b), the conducting in-gap states still exist at both edges, though the two edges are also separated by one-unit-cell distance. For a trivial edge state, it can propagate on both top and lower terraces. The edge states from two nearby edges can couple and induce interference in the area of the groove, but this is not what we observe. While for the case of a 2D topological insulator, its edge states will decay very sharply into the vacuum (in the groove). Therefore, the topological edge states from the sides of the one-unit-cell wide monolayer groove will not strongly couple. The measured partial gap size at the center of the groove is about 50 meV, which is smaller than the bulk gap (80meV) of that on the 2D monolayer, indicating a much weaker coupling compared with the case of the one-unit-cell wide monolayer strip. Fig. 4(c) displays the differential tunneling conductance mapping of the same region in Fig. 4(a) at the energy (100meV) within the bulk gap. The in-gap states for both edges and their spatial evolution are well visualized. For both edges, edge states decay faster into the vacuum (groove) than into the 2D bulk as shown in Fig. 4(d). It should be pointed out that the actual decay of the edge states is much faster than that shown in the measurement, because of the relatively poor spatial resolution at the down step side due to the finite dimension of the STM tip.

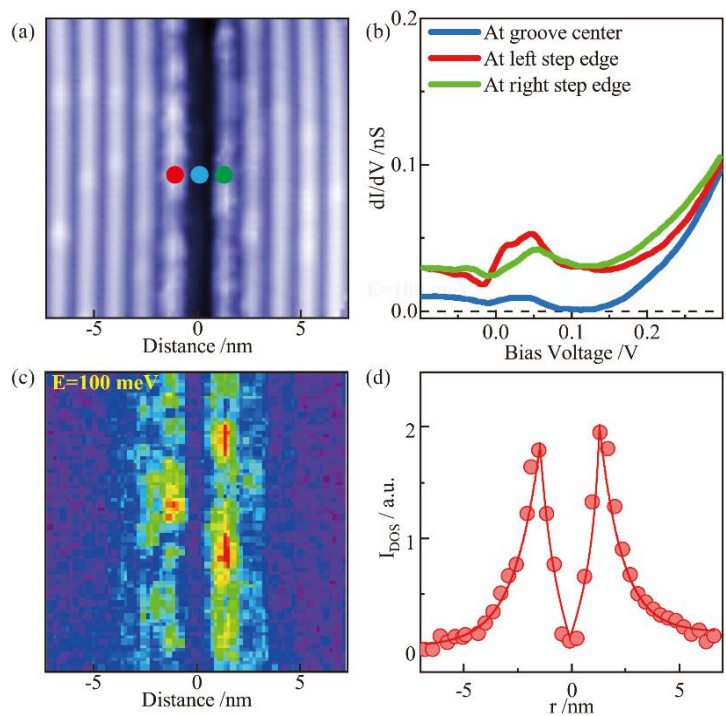

FIG. 4 (a) Topographic image showing two adjacent step edges separated by a one-unit-cell wide monolayer groove (V = 500 mV, I = 0.03 nA). (b) Tunneling spectra taken at the center of the groove (blue) and at the step edges of the groove (red, green), respectively. Their corresponding locations are marked in (a) as colored dots (V = 500 mV, I = 0.5 nA). (c) Corresponding tunneling conductance map at E = 100 meV (Junction set: V= 500 mV, I= 0.3 nA). (d) Integrated DOS within the gap along a line perpendicular to the two edges, showing the faster decay of the edge states into the groove than into the than into the 2D bulk.

In conclusion, our study of HfTe$_5$ single crystals clearly demonstrates that the HfTe$_5$ surface-monolayer has a full gap of ~80 meV and almost constant in-gap states at its edges. These observations are consistent with the theoretical predictions that the bulk of HfTe$_5$ is a weak 3D topological insulator, and that the surface monolayer of HfTe$_5$ is a 2D topological insulator. We also observe the distinct behavior of the in-gap states at two edges separated by one-unit-cell distance. They disappear at both edges of a unit-cell-wide strip and persist at both edges of a unit-cell-wide groove. This remarkable observation, though is not a direct proof, strongly reinforces the topological arguments and undermines the trivial interpretations. Further study of the interaction of the edge states as a function of separation distance would greatly enhance the comprehensive understanding of the topological nature and provide in-depth knowledge for applications.

We sincerely thank Dr. Hong-Ming Weng(翁洪明), Dr Xi Dai(戴希) and Dr. Si-Min Nie(聂思敏) for in-depth discussions. This work is supported by the Chinese Academy of Sciences, NSFC (11227903), BM-STC (Z181100004218007), the Ministry of Science and Technology of China (2015CB921300, 2015CB921304, 2017YFA0302903), the Strategic Priority Research Program B (XDB04040300, XDB07000000), and Beijing Municipal Science & Technology Commission (Z181100004218007.